\documentclass[12pt]{article}

\parskip=\medskipamount            
\def\7#1#2{\mathop{\null#2}\limits^{#1}}        

\def\beee{\begin{equation}}
\def\eeee{\end{equation}}

\begin{document}
\begin{center}
\textbf{OBSERVABLE EFFECTS OF NONCOMMUTATIVE SPACETIME ON THE HYDROGEN ATOM}\\
[5mm]
Benjamin Katsuo Johnson\footnote{email address: katsuo@umd.edu.}\\
Department of Physics\\
University of Maryland\\
College Park, Maryland~~20742\\
University of Maryland Preprint: PP 06-001
\end{center}

\bibliographystyle{unsrt}

\begin{abstract}

We present a brief historical introduction to the motivations behind quantum mechanics and quantum field theory on noncommutative spacetime and provide an insightful technique, readily accessible to the undergraduate student, to examine the measurable effects of noncommutative spacetime on the familiar hydrogen atom.  Finally, we compare our results to those derived from more sophisticated approaches.

\end{abstract}

\section{Introduction}
Noncommutativity is a fundamental concept underlying quantum physics.  All conjugate variables fail to commute and indeed the uncertainty principle itself, 
\beee
(\Delta x)(\Delta p) \geq \frac{\hbar}{2}
\eeee
is based on the commutation relation of position and momentum,
\beee
[x, p] \equiv xp - px = i\hbar.
\eeee

Quantum field theory on noncommutative spacetime is a modification of quantum field theory based on the conjecture that spacetime coordinates at the Planck scale do not commute.  Traditionally, it is assumed that spacetime coordinates are commutative,
\beee
[{x}^{\mu}, {x}^{\nu}] = 0,
\eeee
whereas noncommutative spacetime posits that the commutator of the spacetime coordinates does not vanish.  Many authors assume
\beee
[{x}^{\mu}, {x}^{\nu}]=i\theta^{\mu \nu},
\eeee
where $\theta^{\mu \nu}$ is a fixed antisymmetric numerical matrix; while others assume that all the commutators involving spacetime coordinates and momenta do not vanish.  M.R. Douglas and N.A. Nekrasov$^{\cite{dou}}$ and R.J. Szabo$^{\cite{sza}}$ give reviews of this subject. 

W. Heisenberg first proposed the use of noncommuting spacetime coordinates to reconcile singularity problems connected with ultraviolet divergences in the early days of quantum electrodynamics.  The hope was that smearing out the spacetime coordinates would remove the ultraviolet divergences of quantum field theory associated with very short distances in spacetime.  H. Snyder$^{\cite{sny}}$ formalized and published the first paper on the subject, but it was largely overlooked both because it did not solve the divergence problem and because of the the contemporaneous success of renormalization techniques.  Our approach stems from an argument proposed by S. Doplicher, et al:$^{\cite{dop}}$ a high momentum probe must be employed to measure the position of a particle with great precision.  Since momentum and energy are dependent on the same quantities, the process would create a system with enough energy to bring about gravitational collapse, forming a black hole.  This creates a minimal observable length, deforming the traditional uncertainty relation by adding  an extra term,
\beee
(\Delta x)(\Delta p) \geq \frac{\hbar}{2} + \phi(\Delta p)^2,
\label{deform}
\eeee
where $\hbar$ is the reduced Planck's constant, $h/2\pi$,
\beee
\phi = \frac{(\lambda a_{o})^2}{\hbar},
\label{lambda}
\eeee
where $\lambda$ is a dimensionless parameter whose importance will be revealed in the following section and $a_{o}$ is the Bohr radius.  We will use Eq.~(\ref{deform}) to examine the effects of noncommutative spacetime on the hydrogen atom.  More recently, string theoretic considerations by N. Seiberg and E. Witten$^{\cite{sei}}$ revived interest the subject, but their discussion is beyond the scope of this paper.

\section{The Traditional Hydrogen Ground State}
We begin by following an argument from \textit{The Feynman Lectures on Physics}$^{\cite{fey}}$ employing the uncertainty principle to approximate the size of the hydrogen atom.  From a classical perspective, the attractive electromagnetic force between the proton and electron in a hydrogen atom would cause the electron to radiate all of its energy in the form of light and come crashing down into the nucleus.  In quantum theory, this cannot happen because it would violate the uncertainty principle, as we would be able to accurately determine both the position and momentum of the electron.  Thus we can assume the electron will orbit the nucleus, having an uncertainty in position $(\Delta x)$.  From the uncertainty principle, we can see that the uncertainty in the electron's momentum is $(\Delta p) \approx \hbar/(\Delta x)$.  We can then find the energy of the electron by summing its kinetic and potential energies,
\beee
E = \frac{(\Delta p)^2}{2m}-\frac{1}{4\pi\epsilon_{o}}\frac{e^2}{(\Delta x)},
\label{start}
\eeee
where $m$ is the mass of the electron, $\pi$ is the ratio of a circle's circumference to its diameter, $\epsilon_{o}$ is the permittivity of free space and and $e$ is the fundamental charge, and use the uncertainty principle to replace $(\Delta x)$,
\beee
E = \frac{(\Delta p)^2}{2m}-\frac{1}{4\pi\epsilon_{o}}\frac{e^2(\Delta p)}{\hbar}.
\eeee
We can set $d E/d (\Delta p) = 0$ to find the minimum energy and ground state radius of the hydrogen atom.  This gives us
\beee
(\Delta p) = me^2/(4\pi\epsilon_{o})\hbar
\eeee
and
\beee
(\Delta x) = (4\pi\epsilon_{o})\hbar^2/me^2 \equiv a_{o} = 0.529 \AA,
\eeee
which is the Bohr radius.  Using these to solve for the ground state energy, we find
\beee
E = -\left(\frac{1}{4\pi\epsilon_{o}}\right)^2\frac{m^2e^4}{2\hbar^2}
\eeee
which computes to -13.6eV, the measured ionization energy of the hydrogen atom.
\section{The Perturbed Hydrogen Ground State}
We can now use this approach with the deformed uncertainty relation to determine the effect of noncommutative spacetime on the hydrogen atom.  We begin with the modified commutation relation given by S. Benczik, et al,$^{\cite{ben}}$
\beee
[x, p] = i\hbar(1+\beta p^2),
\eeee
where $\beta = \phi/\hbar$ , which leads to the modified uncertainty relation
\beee
(\Delta x)(\Delta p) \approx {\hbar} + \phi(\Delta p)^2.
\eeee
To find the minimum uncertainty we take
\beee
(\Delta x) \approx (\hbar + \phi(\Delta p)^2)/(\Delta p)
\eeee
and substitute it into Eq.~(\ref{start}):
\beee
E = \frac{(\Delta p)^2}{2m}-\frac{1}{4\pi\epsilon_{o}}\frac{e^2(\Delta p)}{\hbar+\phi(\Delta p)^2}.
\eeee 
We can rearrange this equation algebraically to make it more manageable:
\beee
E = \frac{\hbar(\Delta p)^2 + \phi(\Delta p)^4 -2me^2(\Delta p)/4\pi\epsilon_{o}}{2m\hbar + 2m\phi(\Delta p)^2}.
\eeee
This makes it much easier to take $d E/d (\Delta p) = 0$ and solve for $(\Delta p)$, as we can divide out the denominator after taking the derivative using the quotient rule.  Even then, we are left with a quintic equation in $(\Delta p)$,
\beee
0 = -\frac{1}{4\pi\epsilon_{o}}4m^2\hbar e^2  + 4m\hbar^2(\Delta p) + \frac{1}{4\pi\epsilon_{o}}4m^2e^2\phi(\Delta p)^2+8m\hbar\phi(\Delta p)^3+ 4m\phi^2(\Delta p)^5.
\eeee 
However, assuming $\phi$ is small, (an assumption that will be justified in the following section) we can find a solution with a first order pertubation approximation.  Solving for the first order pertubation, we obtain
\beee
(\Delta p) = \frac{\hbar}{a_{o}}(1-3 \lambda^2)
\eeee
and
\beee
(\Delta r) = a_{o}(1+4\lambda^2).
\eeee
Substituting these values into Eq.~(\ref{start}) gives us
\beee
E = -\frac{1}{4\pi\epsilon_{o}}\frac{e^2}{2 a_{o}} + \frac{1}{4\pi\epsilon_{o}}\frac{\lambda^2 e^2}{a_{o}}.
\label{final}
\eeee
Here we see the importance of $\lambda$ from Eq.~(\ref{lambda}).  The terms of the equation's right hand side have the same variables, except that the second has $\lambda$ which acts as a dimensionless scaling variable.  Since $\lambda$ is squared, the second term is always positive and thus non-commutative spacetime would increase the radius and ground state energy of the hydrogen atom.
\section{Agreement with Literature}
We can compare our result to recent work by Benczik, et al and F. Brau,$^{\cite{bra}}$ who solve the Schr\"{o}dinger equation to calculate the perturbed hydrogen spectrum.  Brau gives an explicit formula for the perturbed hydrogen spectrum,
\beee
E_{nl} = -\frac{m \alpha^2 c^2}{2 n^2}+\frac{\lambda^2 a_{o}^2 m^3 \alpha^4 c^4}{5 \hbar^2}\frac{(4 n - 3(l+\frac{1}{2}))}{n^4(l+\frac{1}{2})}, \label{fbrau}
\eeee
where $\alpha$ is the fine structure constant, $e^2/(4\pi\epsilon_{o})\hbar c$, $c$ is the speed of light, $n$ is the principal quantum number, and $l$ is the angular momentum quantum number.  Substituting $e^2/(4\pi\epsilon_{o})\hbar c$ for the fine structure constant and ground state values $n = 1$ and $l = 0$, Eq.~(\ref{fbrau}) simplifies to
\beee
E = -\frac{1}{4\pi\epsilon_{o}}\frac{e^2}{2 a_{o}} + \frac{1}{4\pi\epsilon_{o}}\frac{\lambda^2 e^2}{a_{o}},
\eeee
which is exactly the same as Eq.~(\ref{final})!  Moreover, Brau does an order of magnitude calculation using the finite size of the electron to put an upper bound on $\lambda \leq 2.83 \times 10^{-7}$ validating our assumption that $\phi$ is small and pertubation theory could be employed with accuracy.  The most sophisticated mathematical tools in our approach were a partial derivative and first order pertubation approximation, both of which can be learned by a typical undergraduate student, yet our results matched very well those derived through much more rigorous means.  We can see from our calculations that noncommuting spacetime coordinates would increase the size of the hydrogen atom very slightly and can expect that more accurate experimental data will allow us to decrease the upper bound or observe a non-vanishing minimal length.

\section*{Acknowledgements}
We thank Professor O.W. Greenberg for his apt tutelage and infinite patience.

\end{document}